# Virtual Machine Consolidation for Datacenter Energy Improvement

Sina Esfandiarpoor [a], Ali Pahlavan [b], Maziar Goudarzi [a,b]

[a] *Energy Aware System (EASY) Laboratory, Computer Engineering Department, Sharif University of Technology, Tehran, Iran (Kish Campus)*

[b] *Energy Aware System (EASY) Laboratory, Computer Engineering Department, Sharif University of Technology, Tehran, Iran*

**Abstract**

Rapid growth and proliferation of cloud computing services around the world has increased the necessity and significance of improving the energy efficiency of could implementations. Virtual machines (VM) comprise the backend of most, if not all, cloud computing services. Several VMs are often consolidated on a physical machine to better utilize its resources. We take into account the cooling and network structure of the datacenter hosting the physical machines when consolidating the VMs so that fewer racks and routers are employed, without compromising the service-level agreements, so that unused routing and cooling equipment can be turned off to reduce energy consumption. Our experimental results on four benchmarks shows that our technique improves energy consumption of servers, network equipment, and cooling systems by 2.5%, 18.8%, and 28.2% respectively, resulting in a total of 14.7% improvement on average in the entire datacenter.

*Keywords:* Cloud computing, Energy efficiency, Consolidation, Datacenter, Service-level Agreements

## 1. Introduction

Cloud computing has been recently brought into focus in both academic and industrial community due to increasing pervasive applications and consequently resource allocation over the Internet. Moreover, users can access the cloud services anytime and anywhere. To this end, cloud services are referred to three categories: Infrastructure as a Service (IaaS), Platform as a Service (PaaS) and Software as a Service (SaaS). With growing these services, most of companies such as Google, Yahoo, Microsoft and IBM extend their datacenters. Therefore, increasing number of clusters and servers in datacenters leads to higher energy consumption costs.

The high energy consumption of datacenter has made it inevitable to move toward designing and deployment of energy-efficient techniques for building a green datacenter. In recent years, many efforts have been made to improve the energy efficiency of virtualized datacenter from different aspects including processor, storage and, network energy management. Moreover, Visualization is one of the important techniques to reduce energy consumption of datacenters. In this technique, virtual machines (VMs) are assigned to minimum number of physical machines such that the utilization of turned on physical machines is also maximized.

In this work, we consider requested MIPS of each VM instead of CPU load as a criterion in ranking VMs for placing on the hosts. For example, assume that a company has three servers (S1-S3) and each server has one CPU core with performance equivalent to 1000 Million Instructions Per Second (MIPS). There are five VMs (VM1-VM5) with the performance of 250, 500, 1000, 750 and, 500 MIPS respectively. Suppose that all the servers are initially idle and the first incoming users request needs 100, 200, 600, 300 and, 200 MIPS of each VM respectively to run. There are two possible solutions to assign VMs to servers in this case.

**Solution 1:** A VM-based assignment may choose the solution leading to assign VM1 and VM4 to S1, VM2 and VM5 to S2 and finally, VM3 to S3. Therefore, we use three servers with utilization of 40%, 40% and 60% respectively.

**Solution 2:** A Request-Based assignment may lead to assign VM1 and VM3 to S1 and the remaining VMs to S2, and consequently, the S3 can be turned off. Therefore, we use two servers with utilization of 70%.

In the first solution, Service Level Agreement (SLA) will not be violated but, the utilization of servers is low and the number of active servers is higher than the second solution leading to higher energy consumption. In the second solution, SLA may be violated due to variation of requests MIPS at any time. But, this solution increases the utilization of ON servers resulting in an efficient use of datacenter resources. To avoid SLA violation, an upper bound threshold value is considered for utilization of each server in such a way that some of the VMs which are run on a server have to migrate to other physical machines when the utilization of this server becomes more than this threshold value.

To the best of our knowledge, this paper considers four energy-aware resource management algorithms for virtualized datacenters so that, total energy consumption of datacenters is minimized. According to proposed VM placement methods, cloud computing can be a more sustainable to move forward for future generations.

With an appropriate assignment, we can decrease the number of active servers. This problem is similar to the bin packing problem. Bin packing problem is an NP-hard problem so that, approximation algorithm is used to solve bin packing problem. The most well-known methods to assign VMs to physical machines are First Fit Decreasing, Round Robin and Best Fit Decreasing. In First Fit Decreasing, the VMs are first sorted in decreasing order of their sizes, and then each VM is inserted into the first server of the list of server with enough resources. The advantage of this method is its simplicity. But this method cannot find the most appropriate server so, it is not energy aware. Round Robin is like First Fit, however in this method VMs are assigned to servers uniformly. Thereby, it avoids servers overloading, and network traffic is balanced but the disadvantage is that no server or network switch is left idle so none of them can be turned off to save power. The last method is best fit decreasing, in this method after sorting VMs in decreasing order of their size, allocate VM to the most sufficient server. This method performs better placement rather than others.

In this paper, we first propose a VM placement algorithm that improves the modified best fit decreasing (MBFD) algorithm which is presented in [1]. Then, we present three other VM placement methods to consolidate VMs in the servers to minimize number of active racks. This in turn leads to turn off cooling systems, network switches of idle racks in order to reduce total energy consumption of datacenter. These algorithms are: Place VMs Rack by Rack (RBR), Place VMs in None-Underutilized Rack (NUR), and, Only Migrate Underutilization Racks (OMUR). In summary, our contributions in this work are described as follows:

- We evaluate our proposed VM placement algorithms under a typical virtualized datacenter comprising rack, cooling structure and network topology.
- We merge the VMs of new arrived requests and overloaded VMs. This will ensure that VMs are assigned to minimum number of ON servers.
- We also consider the utilization of racks in our VMs placement algorithms. This leads to improve energy consumption of cooling and network of datacenter.
- After the determining location of incoming and overloaded VMs, We also try to place VMs of underutilized racks on the non-underutilized racks to turn off servers of underutilized racks.

We compare our proposed algorithms to the best conventional method so that results show up to 14.7% energy improvement at a common datacenter.

The rest of paper is organized as follows. In section 2, we explain the related work and some VM placement algorithms. Section 3 introduces used network topology, server and cooling system energy model to calculate total energy consumption of datacenter via a mathematical formulation. In the next section, we propose our VMs placement algorithms. Section 5 is related to simulation environment and evaluated results. Finally, we draw our conclusion and talk about future work in the last section.

## 2. Related Work

In recent years, several techniques have been proposed on VMs placement techniques in virtualized datacenters. The authors in [2] have studied round robin algorithm to schedule and consolidate VMs. They have proposed a new strategy for VMs placement and migration that is called Dynamic Round-Robin (DRR). DDR as the extension of the Round-Robin method tries to reduce the number of active physical machines using two rules. In the first rule, if the running of a VM on a server has finished and there are still other VMs on the same physical machine, this physical machine will not accept new VMs. In the second rule, if a physical machine remains in the first rule for a sufficiently long period of time, instead of waiting for the VMs to finish, the physical machine will be forced to migrate the rest of its VMs to other physical machines which in turn leads to shut down physical machine after the migration completion.

Wang Xiaoli et al. [3] have improved Bin-Packing algorithm. They considered a threshold to avoid inopportune VM migration. To this end, if sum of VMs resources of a server becomes less than this threshold value, the VMs of this server are migrated to other servers with enough resources to turn off this server. Note that this approach does not consider the cost of frequently swapping of VMs under migration. Also, they do not minimize the potential increase of the server's utilization to prevent SLA violations when the utilization of servers increases by VMs.

In [4], a new VM placement and migration method is presented for data-intensive applications in cloud computing. In this work, the goal is to reduce time of data transfer in order to increase total performance. They assigned VMs to servers having the minimum access time to data. In their migration approach, an execution time threshold for each application is defined in order to complete processing its related data. If data access times of each VM be larger than its threshold value, VM is migrated to another server with minimum data access time. For placement of new arrived VM, first data then new arrived VM is allocated in this host.

The authors in [5] introduced a way to reduce energy consumption of datacenter by determining best candidate of VM for migration and finding best physical machine having enough resources to accept this VM. In live migration [6], pre-copy of data from source physical machine to destination physical machine occupies 80 percentage of migration period. This part is CPU and network intensive that consumes the most energy. To decrease the cost of VMs migration, the servers are selected which have smaller memory and larger CPU reservation. They also investigate co-migration effect in such a way that some VMs want to place on some physical machine concurrently. The authors solved this problem with a heuristic algorithm to find the order of VM for migration.

To this end, available memory, network and disk resources of physical machine and demanded memory, network and disk space of VM are considered as the fitness function parameters of heuristic algorithm. However, this method is inefficient in online power management due to high computational overhead.

Anton Beloglazov et al. [1] proposed an architectural framework and principles for energy efficient cloud computing. Indeed, the authors investigated a principle of architecture for energy efficient management of cloud systems and energy efficient resource allocation policies considering characteristics of device`s power usage based on this architecture. The authors modified Best First Deceasing algorithm to improve energy efficiency of the virtualized datacenter. In this purposed algorithm, the VMs are first sorted in decreasing order of their utilizations and then these VMs are allocated to the hosts having minimum increment of energy consumption.

The authors defined three double-thresholds policies for VMs selection. To this end, VMs are needed to migrate when the CPU utilization of hosts exceeds the upper threshold value to prevent SLA violations and also, all of VMs on the hosts should be migrated when the CPU utilization of hosts falls below lower threshold to turn off idle hosts. The goal of first VM selection policy, The Minimization of Migrations Policy, is to minimize the number of VMs needed to migrate from a host. In this way, for choosing the best VM two conditions should be considered: First condition, the VM is chosen that has the most difference between the host's utilization and the upper threshold. Second condition, if the VM needs to be migrated from the host, the difference between the upper threshold and the new utilization is the minimum across the values provided by all the VMs. The Second policy is the highest potential growth policy. In this method, the VM with most potential of CPU usage increasing is selected. The last policy is The Random Choice Policy that selects VMs randomly until CPU utilization of related host falls below upper threshold.

They could improve energy consumption of datacenter rather than previous work. This approach tries to use minimum number of hosts to reduce energy consumption datacenter but, energy consumption of a typical datacenter including servers, network and cooling system is not considered in this work. In our work, we try to allocate VMs to the servers which are placed in racks to improve total energy consumption of datacenter by turning off all idle servers, switches and cooling systems.

## 3. System model

In this section, we first introduce used datacenter configuration and then, we model the total power consumption of a datacenter with three components: servers, cooling system and network switches power consumption that are defined as follows.

### 3.1 Datacenter configuration

A datacenter comprises a structure typically as: Network Switches, Cooling System, Racks and Servers. Fig. 1 depicts a datacenter configuration with eight racks. Each rack contains several servers so that, each server has a dedicated power unit and a cooling fan.

The used datacenter has several rack-based cooling systems that are dedicated to each rack. In this structure, the cold air is delivered directly inside the rack and heat generated by the servers is transferred through heat risers to cooling rack. Hence, the efficiency of cooling system increases since the exact location of the air conditioner to the target load is determined.

The used network topology in the most of datacenters is a tree structure; this is shown in Fig. 1. In this case, several top of rack switches connect to one aggregate switch and, in the higher level, the aggregation-layer switches connect to a core router in order to track and route the migrated VMs from one sever to another server in another location.

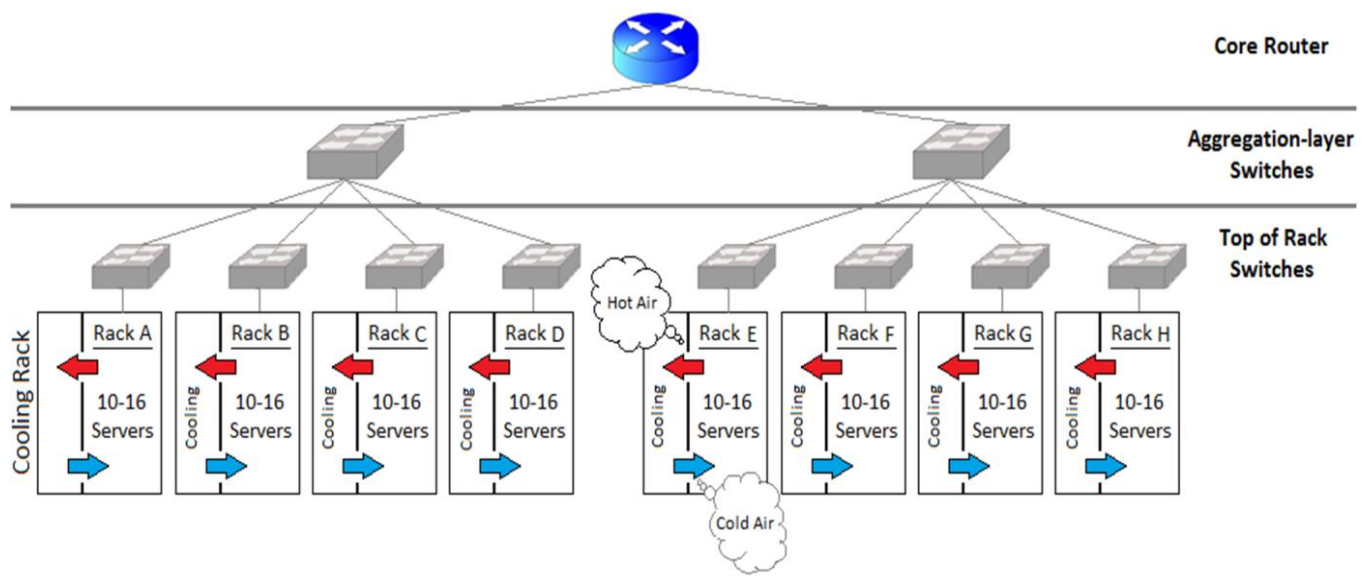

Fig. 1 Datacenter configuration: Location of Servers, Cooling systems and Network topology.

## 3.2 Power model

In this work, we model the total power consumption of a datacenter with three components: servers, cooling device and network power consumption that are defined as follows.

The major components of server that consume power are CPU, memory, storage and network interface in which, CPU has the most effect in power consumption of server. Previous investigations [7-10] have shown that the power-to-frequency relationship in a server is linear for CPU-intensive workloads. Also, these studies have shown that an idle server consumes on average about 70% of power consumption of a full utilized server. Hence, the power model of servers can be calculated similar to [1] as:

$$P_i = K.P_{imax} + (1-K).P_{imax}.U_i \qquad (1)$$

where, $P_i$ denotes power consumption of $i^{th}$ server. $K$ shows fraction of power consumption of an idle server independent of its utilization. In this paper, typical K is 70% [1]. $P_{imax}$ and $U_i$ represent maximum power consumption and utilization of $i^{th}$ server respectively.

According to workload variability, CPU utilization may change over the time. Therefore, $U(t)$ is defined as the utilization of server in time t. To this end, energy consumption of each server in a period of time [$t_1$,$t_2$] will be calculated as:

$$E = \int_{t_1}^{t_2} P(U(t)).dt \qquad (2)$$

Total power consumption of datacenter contains consumed computation power by servers (Eq. (1)), cooling system and, network components. To this end, the summation of power of all ON switches in each layer of network topology forms the power consumption of related layer. Therefore, the consumed network power ($P_{NET}$) is expressed as follows:

$$P_{NET} = P_{ToR} + P_{Agr} + P_{CR} \qquad (3)$$

where $P_{ToR}$, $P_{Agr}$ and $P_{CR}$ denote power consumption of top of rack switches, aggregation-layer switches and core router respectively. For total cooling power consumption ($P_C$), if $j^{th}$ rack is turned on, the related cooling rack will be turned on and consumes $P_j$ power. Therefore, $P_C$ is defined as $P_C = \sum_{j=1}^{N_r} P_j$ where $N_r$ determines total number of racks. Finally, the total power consumption of datacenter $P_{DC}$ is defined as follows:

$$P_{DC} = P_S + P_C + P_{NET} \qquad (4)$$

where $P_S = \sum_{i=1}^{N_s} P_i$ represents total power consumption of servers and $N_s$ specifies the total number of servers in all racks.

## 4. Proposed Energy-Aware Virtual Machine Placement Algorithms

### 4.1 Problem Statement

The flow diagram of our infrastructure is shown in Fig. 2. In each epoch, we choose the appropriate VMs from overloaded servers and then, merge them with new arrived VMs as inputs for VMs placement algorithm block at the first step. We will explain our VMs placement methods later in this section. Ultimately, after placing these VMs on the servers, we find all VMs of underutilized servers to migrate into the best other servers using VMs placement algorithm. This fact leads to switching idle servers to the sleep mode to reduce the total power consumption of datacenter.

For VMs migration, we consider two thresholds comprising lower and upper bound CPU utilization thresholds so that, we keep the utilization of CPUs between these thresholds. If the CPU utilization becomes less than the lower threshold value, all VMs have to be migrated from this host and the host has to be switched to the sleep mode in order to eliminate the idle power consumption. If the utilization exceeds the upper threshold, some VMs have to be migrated from the host to reduce the utilization. The aim is to preserve free resources in order to prevent SLA violations due to the consolidation in cases when the utilization by VMs increases.

Similarly to [1], for choosing the best VM to migrate from the host, two conditions should be considered. First, the VM is chosen that has the most difference between the host's utilization and the upper threshold. Second, if the VM needs to be migrated from the host, the VM should provide the minimum difference between the upper threshold and the new utilization in presence of all the VMs. Otherwise, if there is no such a VM for migration, the VM with the highest requested MIPS is selected. The algorithm continues until the utilization of the host becomes less than the upper utilization threshold. This policy leads to the minimum number of VMs migration in order to reduce the migration overhead.

In the following sections, we design and implement several VMs placement algorithms to improve power consumption of datacenter.

### 4.2 OBFD algorithm

We modified the MBFD algorithm in [1] to improve VMs placement. In proposed algorithm (OBFD), we first sort the VMs in decreasing order of their required MIPS instead of current CPU utilizations. After ranking VMs, we try to find the best server for each VM that leads to the minimum increasing of power consumption of datacenter. Therefore, in this step, we find the best server among all non-underutilized and non-empty servers; if no server could be found for assigning the VM, the algorithm tries to find the best server among all underutilized servers and

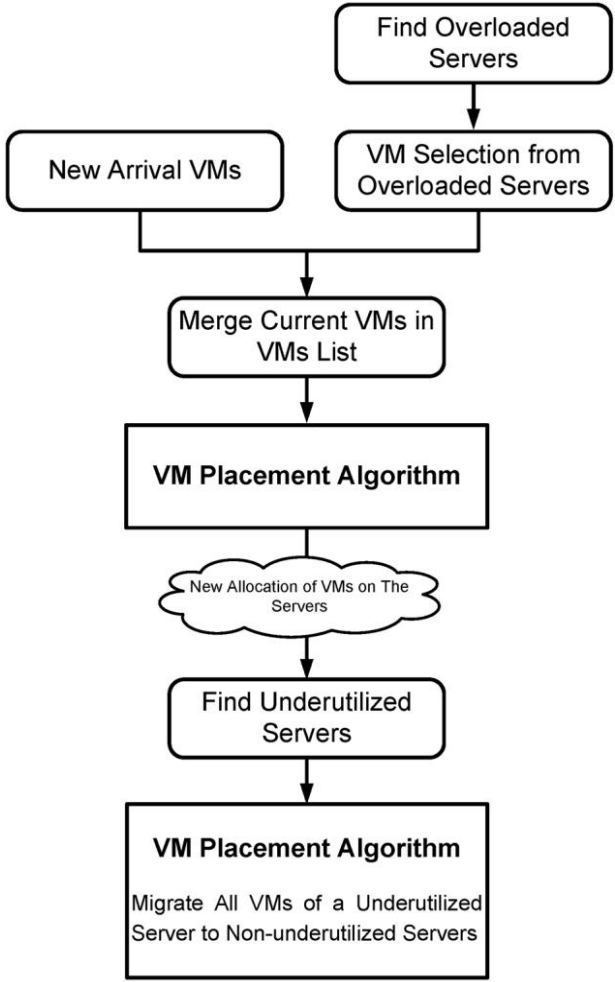

Fig. 2 The high-level system infrastructure.

**Algorithm 1**: Our Modified Best Fit Deceasing (OBFD)

**Inputs**: ServersList, VMsList
**Output**: AllocationOfVMs
1. **Initialize**: EmptyServersQueue, UnderutilizaedServersQueue
2. SortVMsInDecreasingMIPS()
3. **foreach** VM in VMsList **do**
4.    MinPower ← *MAX*
5.    AllocatedServers ← *NULL*
6.    **foreach** Server in ServersList **do**
7.      **if** Server is empty **then**
8.        EmptyServersQueue ← Server
9.      **else**
10.        **if** Server is a underutilized server **then**
11.          UnderutilizaedServersQueue ← Server
12.        **else**
13.          **if** Server has enough resources **then**
14.            Power ← EstimatedPower by Eq. (1)
15.            **if** Power < MinPower **then**
16.               AllocatedServers ← Server
17.               MinPower ← Power
18.    **if** AllocatedServers = *NULL* **then**
19.      **foreach** Server in UnderutilizaedServersQueue
20.        **do** lines 13-17
21.    **if** AllocatedServers = *NULL* **then**
22.      **foreach** Server in EmptyServersQueue
23.        **do** lines 13-17
24.    **if** AllocatedServers ≠ *NULL* **then**
25.      Allocate VM to AllocatedServers
26. **return** **AllocationOfVMs**

finally, if it could not find among all ON servers, the algorithm turns on a server from the empty servers list to place VM on it. The pseudo-code for the algorithm is presented in Algorithm 1. This algorithm guarantees the minimum number of ON severs to reduce power consumption of servers whereas these active servers do not place on the minimum number of racks to reduce cooling system and network power consumption.

## 4.3 Placement Algorithm with considering rack utilization

In this section, we consider the rack utilization for consolidate VMs due to turn off the cooling system and related switches of idle racks. We also consider a lower bound threshold for racks utilizations. To this end, if a rack utilization becomes less than this threshold, entire VMs of this rack are migrated to other racks so that, this leads to turn off switches and cooling units of this rack.

We present three rack-aware policies for VMs placement: place VMs Rack By Rack, Place VM in None-Underutilized Rack, Only Migrate under-utilized Rack. In the following section we explain our proposed policies.

### 4.3.1 Place VMs Rack by Rack (RBR)

In this policy, we first sort racks and current VMs in decreasing order of utilization and required MIPS respectively. Then, we try to find the first rack having enough resources from the list of sorted racks in order to assign first VM from the list of sorted VMs to selected rack. Finally, for choosing the best server of selected rack, the OBFD will be run to complete the placement of this VM. Algorithm 2 shows the pseudo code of the proposed VM placement.

In this method, for placing each VM on the servers of determined rack, if the active servers do not have enough resources, a sleep (turned off) server will be turned on in this rack. After assignment of both current and overloaded VMs groups, we try to migrate the VMs of underutilized racks having utilization less than lower bound threshold to servers of other non-underutilized racks having enough resources.

**Algorithm 2**: Place VMs Rack by Rack (RBR)

**Inputs**: RacksList, VMsList, ServersList
**Output**: AllocationOfVMs
1. SortVMsInDecreasingMIPS()
2. SortRacksInDecreasingUtilization()
3. **foreach** VM in VMsList **do**
4. | SelectedRack ← FirstRackfromRackListWithEnoughResources()
5. | **foreach** Server in SelectedRack **do**
6. | | AllocatedServers ← FindAppropriateServerbyOBFD()
7. | **if** AllocatedServers ≠ *NULL* **then**
8. | | Allocate VM to AllocatedServers
9. return **AllocationOfVMs**

### 4.3.2 Place VMs in None-Underutilized Rack (NUR)

In this policy, we try to find the best server among all non-underutilized racks for each VM in sorted VM list using OBFD algorithm. If OBFD could not find a server in the none-underutilized racks, the algorithm tries to find an appropriate server in underutilized racks. Similar to RBR policy, the VMs of underutilized racks will be migrated to other racks after placing all current and overloaded VMs groups on hosts.

In this technique which is presented in Algorithm 3, we search among all none-underutilized racks to find the appropriate servers for VMs. This provides the efficient use of active servers and in turn, gives the opportunity to prevent turning on extra servers as much as possible. Therefore, this algorithm leads to a decrease in energy consumptions of servers due to using the minimum number of ON servers.

**Algorithm 3**: Place VMs in None-Underutilized Rack (NUR)

**Inputs**: RacksList, VMsList, ServersList
**Output**: AllocationOfVMs
1. SortVMsInDecreasingMIPS()
2. **foreach** VM in VMsList **do**
3. | **foreach** Server in None-UnderutilizedRacks **do**
4. | | AllocatedServers ← FindAppropriateServerbyOBFD()
5. | **if** AllocatedServers = *NULL* **then**
6. | | **foreach** Server in UnderutilizedRacks **do**
7. | | | AllocatedServers ← FindAppropriateServerbyOBFD()
8. | **if** AllocatedServers ≠ *NULL* **then**
9. | | Allocate VM to AllocatedServers
10. return **AllocationOfVMs**

### 4.3.3 Only Migrate Under-utilized Racks (OMUR)

In this policy, we use rack consolidation technique along with the VMs placement to reduce datacenter energy consumption. This will ensure that incoming VMs are assigned to minimum number of ON servers in minimum number of racks.

In this process, first VMs are assigned to the active servers in non-underutilized racks so that the difference between the host's new utilization and the upper utilization threshold is minimized. If there is no such a VM, we sort all the racks in decreasing order of utilization and then, the servers in each rack are sorted in decreasing order of their utilizations. Therefore, the underutilized and turned off servers in higher utilized rack are located in top of list. To this end, the VMs are placed on the servers in the racks according to list by running OBFD algorithm so that this leads to the racks reach to its maximum utilization.

After VMs placement, we try to migrate the VMs of underutilized racks to servers of other non-underutilized racks having enough resources. Therefore, the rack consolidation technique leads to reduce total energy consumption of datacenter comprising servers, cooling system and network switches energy consumption. The pseudo-code for the algorithm is shown in Algorithm 4.

**Algorithm 4**: Only Migrate Under-utilized Racks (OMUR)

**Inputs**: RacksList, VMsList, ServersList
**Output**: AllocationOfVMs
11. SortVMsInDecreasingMIPS()
12. **foreach** VM in VMsList **do**
13. | **foreach** Server in None-UnderutilizedRacks **do**
14. | | AllocatedServers ← FindServerWithMinimumUtilizationGap()
15. | **if** AllocatedServers = *NULL* **then**
16. | | **foreach** Server in UnderutilizedRacks **do**
17. | | | AllocatedServers ← FindAppropriateServerbyOBFD()
18. | **if** AllocatedServers ≠ *NULL* **then**
19. | | Allocate VM to AllocatedServers
20. **foreach** Rack in RacksList **do**
21. | Migrate VMs of underutilized Racks
22. return **AllocationOfVMs**

## 5. Simulation and Evaluation

In this section, we provide the experimental results of applying our scheduling methods on the benchmarks. The detailed specification of our architecture and the benchmarks are explained in next section. We investigate an analysis on energy efficiency, SLA violations, number of migrations and run-time of the proposed approaches and then, the results are discussed.

## 5.1 Experiment Setup

The parameters used to model the datacenter structure are as follows:

There are two rows of racks in the structure. Each row consists of four 42U racks. Each rack has ten servers. Each server has single core CPU with performance equivalent to 2000 MIPS, 10 GB memory, 1TB storage and 1 GB network interface. According to this structure, each server consumes 175W in idle state and up to 250W with 100% CPU utilization [1]. Each VM requires one core CPU with performance equivalent to 250, 500, 750 or 1000 MIPS, 128 MB memory, 1 GB storage.

We provide four benchmarks by uniformly distributed to generate random variables of requests in terms of MIPS. To this end, we consider four groups of VMs in each epoch. The first group includes VMs that leave the datacenter in the new epoch. The second group includes VMs whose request arrival MIPS drop in the new epoch and the third group includes VMs whose request arrival MIPS rise in the new epoch. Finally, the fourth group includes new VMs that arrive in each epoch. Similar to [1], each experiment has been performed 4 times with different benchmarks.

We consider a simple three-layered tree structure for network topology in such a way that the available servers in a rack are connected to a top of rack switch. In the next layer, top of rack switches of four racks are connected to one aggregate switch and finally, the aggregate switches are connected to one core router at last layer. The type of top of rack, aggregation switches and core router is HP5920, HP6600 and HP8800 that dissipate 366W, 405W and 3500W respectively [11].

We consider rack-based cooling mechanism to provide cold air for each rack. The type of cooling unit is HP Modular Cooling System G2 that consumes 950W for cooling each rack [11]. According to [1], we choose the best threshold pairs for migration policy. To this end, the lower and upper threshold values are defined 40% and 80% respectively to determine overloaded and underutilized servers and racks.

## 5.2 Simulation results

### 5.2.1 Evaluating the Proposed VM Placement Methods

In this section, we compare the achieved total energy consumption, servers, network and cooling cost results in the case of five different VM placement methods:

- Baseline Modified Best Fit Decreasing (MBFD)
- Our Best Fit Decreasing (OBFD)
- Rack by Rack (RBR)
- None-Underutilized Rack (NUR)
- Only Migrate Under-utilized Racks (OMUR)

In the following figures (Fig. 3-8), servers, network and cooling costs and total energy consumption of datacenter compared to MBFD are exhibited for the mentioned benchmarks.

Fig. 3 shows the average number of active servers which affects energy consumption of servers directly. In this case, Fig. 4 specifies 2.5% power enhancement on average for all benchmarks by OBFD and OMUR. In general, OBFD and OMUR try to reduce the number of ON servers among all available severs in all racks leading to a decrease in servers energy consumption whereas RBR and NUR try to use the minimum number of racks instead of severs to improve the network and cooling cost. Therefore, RBR and NUR lead to an increase in energy consumption of servers compared to MBFD.

One of the most important factors that affect energy consumption of cooling and network units is the number of active racks. As shown in Fig. 5, the OBFD, RBR, NUR and OMUR reduce the number of active racks 1%, 27.9%, 22.6% and 28.2% on average for all benchmarks respectively. This factor affects the energy consumption of network components and cooling system.

The OMUR and RBR algorithms use rack consolidation technique to turn off cooling system and top of rack switches in order to reduce energy consumption of major parts including cooling and network. In Fig. 6 and 7, the results show 18.6%, and 18.8% energy improvement on average for network and 27.9%, and 28.2% improvement on average for cooling part by RBR and OMUR respectively. The NUR method try to place VM on non-underutilized racks in order to prevent turning on more racks leading to 15.1% and 22.6% network and cooling cost improvement on average compared to MBFD. The OBFD algorithm minimized the number of active servers. Note that in this case, randomly distributed VMs over higher number of racks, OBFD fails to reduce the number of ON racks effectively. Therefore, the OBFD improves the energy consumption of network and cooling components 0.7% and 1% on average respectively.

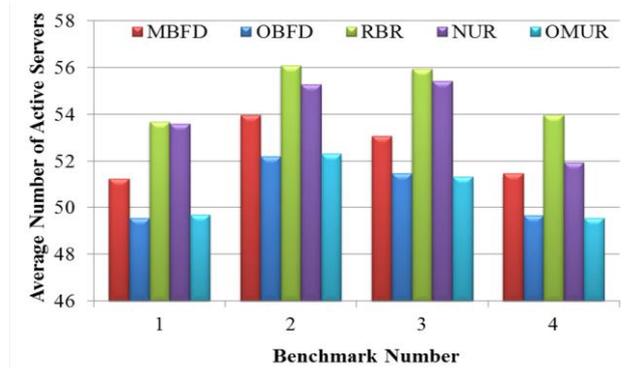

Fig. 3 Average number of active servers for all benchmarks.

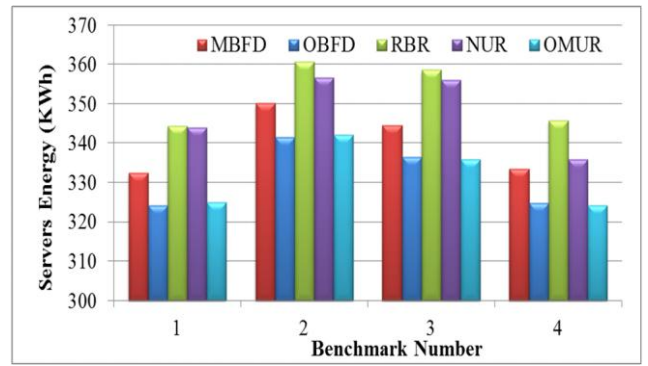

Fig. 4 Energy consumption of servers for all benchmarks.

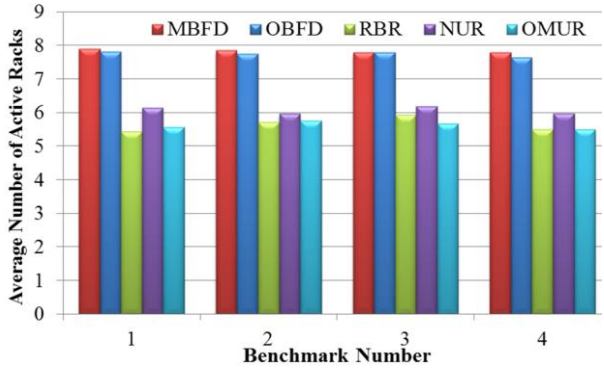

Fig. 5 Average number of active racks for all benchmarks.

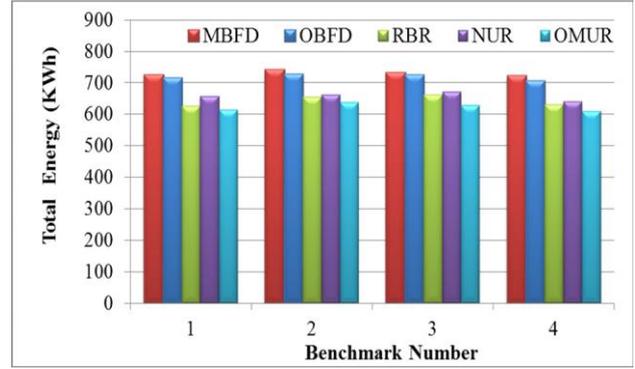

Fig. 8 Total energy consumption of datacenter for all benchmarks.

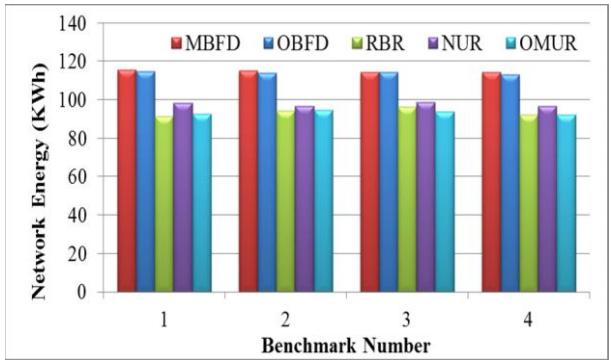

Fig. 6 Energy consumption of network for all benchmarks.

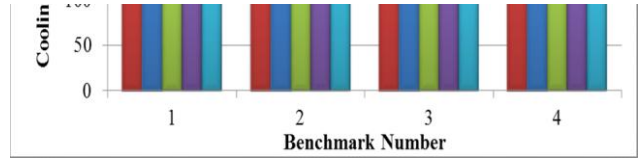

Fig. 7 Energy consumption of cooling system for all benchmarks.

In the following figure (Fig. 8), total energy consumption including server, network and cooling cost, and energy improvement percentage compared to MBFD are shown for mentioned benchmarks. The results show 1.6%, 11.8%, 9.8% and 14.7% energy savings on average for OBFD, RBR, NUR and OMUR respectively compared to the baseline conventional MBFD method. The OMUR algorithm surpasses other proposed algorithms on average since it has the flexibility of choosing the minimum number servers in minimum number of racks to reduce the total energy consumption by turning off idle network switches, cooling systems and servers. In general, since the cooling system is one of the major contributors to energy cost encompassing about 30% or more of total energy cost in a large scale datacenter [12], the proposed rack-based algorithms (RBR, NUR and OMUR) provide high efficiency compared to the OBFD and MBFD according to datacenter configuration (Fig. 1). Also, a typical datacenter usually operates in 20-30% utilization rates [13] therefore, we can achieve overall improvement in datacenter energy consumption under the proposed heuristics because; all racks are not fully turned on at different utilization rates.

### 5.2.2 Evaluating the Migration Cost

In this simulation, the time needed to perform a migration of a VM is calculated as the size of its memory in such a way that the network bandwidth is divided to migrate the VMs. the data of VMs must be stored on a Network Attached Storage (NAS) to prevent copying the VM's storage. Therefore, the performance overhead of migration is low; however, it creates an extra CPU load [14].

In Fig. 9, Total number of migration varies under different benchmarks since this metric is dependent on required CPUs load of VMs in different times. This is due to fact that we use uniform distribution to create all benchmarks. At worst case, we have up to 3.8% and 7.5% the number of migrations increment on average compared to MBFD for OBFD and OMUR respectively. This is due to fact that OBFD and OMUR leads to an increasing in the overloaded servers and underutilized racks for VMs migration. But, we achieve 3.4% and 1% enhancement on average for RBR and NUR for all benchmarks respectively.

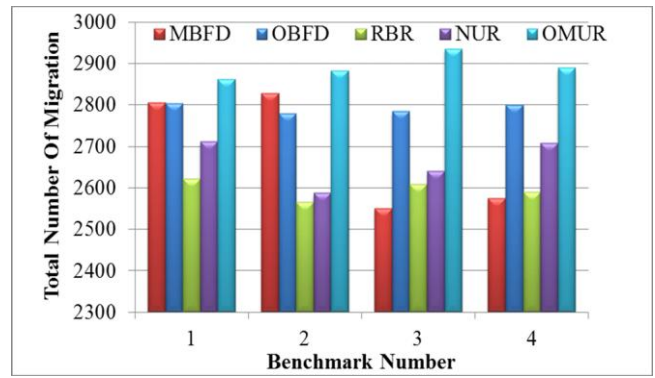

Fig. 9 Total number of migrations for all benchmarks.

### 5.2.3 Evaluating the proposed approaches under SLA

In this work similar to [1], we define SLA violations as the percentage of SLA violation events relatively to the total number of the processed time frames. We define that an SLA violation occurs when a given VM cannot get the amount of MIPS that are requested. This can happen in cases when VMs sharing the same server require a CPU performance that cannot be provided due to the consolidation. This metric shows the level by which the Quality of Service (QoS) requirements negotiated between the resource provider and consumers are violated due to the energy-aware resource management. It is assumed that the provider pays a penalty to the costumers in case of an SLA violation.

As shown in Fig. 10, SLA is dependent on variation of CPUs load in different size and time of arriving VMs. Therefore, the amount of SLA will be changed under different benchmarks. The results show up to 7.1%, 5.4% and 0.4% SLA violations for RBR, NUR, and OMUR respectively. OBFD improves 6.5% In general, The SLA violation of proposed algorithms is tried to keep below the SLA violation of MBFD algorithm so that the SLA violation does not exceed 2.3% for all benchmarks.

### 5.2.4 Run-time of Proposed Algorithms

To measure average run-time of proposed algorithms, we first run each algorithm for placing 3000 VMs on the servers and then, the average run-time of allocating each VM to a server is calculated. The detailed specification of our system for measuring is Intel Core i5 CPU M540 with 6GB memory.

The run-time of placing each VM for OBFD, RBR, NUR and OMUR is 844, 400, 542 and 889 milliseconds (ms) on average respectively. It is clear that than the other algorithms and hence the run-time of OMUR is greater than three other algorithms. RBR has the least run-time because we need to search all servers of a rack for finding the best server.

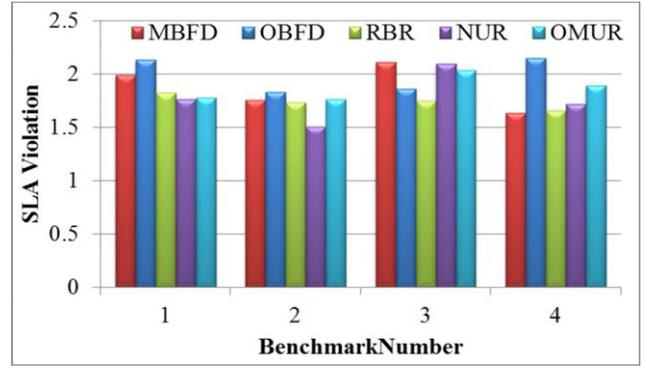

Fig. 10 SLA violations percentage for all benchmarks.

### 5.2.5 Discussion

In this section, we show the energy saving, number of migration and SLA violation improvement of all proposed techniques compared to best conventional method (MBFD) and the results are discussed.

From the presented results (Table 1), we can conclude that the usage of the OMUR algorithm provides the best energy saving with the acceptable SLA violation. But, the number of VM migrations and run-time among the evaluated algorithms leads to high performance overhead. Moreover, the results show the flexibility of the RBR algorithm due to least number of VM migrations and SLA violation with admissible total energy improvement. Also, RBR provides least run-time overhead to allocate a VM to a host on average. In general, the rack-based algorithms improve total energy consumption of datacenter with keeping SLA violations below 2.3%. On the other hand, we

Table 1 Servers, Network and Cooling energy saving percentages on average for all benchmarks.

| Algorithms | Number of Servers and Racks Improvement (%) | | Energy Savings (%) | | | | Number of Migrations Improvement (%) | SLA Violations Improvement (%) |
|---|---|---|---|---|---|---|---|---|
| | Servers | Racks | Servers | Network Components | Cooling System | Total | | |
| OBFD | 3.3 | 1.0 | 2.5 | 0.7 | 1.0 | 1.6 | -3.8 | -6.5 |
| RBR | -4.8 | 27.9 | -3.6 | 18.6 | 27.9 | 11.8 | 3.4 | 7.1 |
| NUR | -3.1 | 22.6 | -2.3 | 15.1 | 22.6 | 9.8 | 1.0 | 5.4 |
| OMUR | 3.3 | 28.2 | 2.5 | 18.8 | 28.2 | 14.7 | -7.5 | 0.4 |

OMUR is more complex

try to reduce the computational overhead including migration and run-time of proposed methods.

## 6. Conclusion and Future Work

In this paper, we proposed VM consolidation algorithms for cloud datacenter energy reduction considering SLA constraint. We introduce a new modified Best Fit Decreasing method (OBFD) and three (RBR, NUR and OMUR) VM placement techniques to find the best location of each VM on the servers based on the typical datacenter configuration. We also used a migration policy to migrate selected VMs of overloaded servers and all VMs of servers in underutilized racks to effectively shut down idle servers and consequently cooling system and network component of idle racks to save energy. Experimental results showed that by using our proposed algorithms, OBFD, RBR, NUR and OMUR, up to 1.6%, 12%, 12.5% and 14.7% energy reduction can be obtained compared to conventional approach (MBFD) respectively so that SLA violation was kept below 2.5%. Finally, we investigated the effectiveness of our approach under the number of migration to prevent increasing performance overhead. Therefore, our algorithms will be even more effective in future growth of cloud computing.

This research work is planned to develop a software platform that supports the energy-efficient management and resource allocation under different datacenters configuration including severs and cooling structure, and network topology. Also, we plan to utilize a thermal model of the datacenter based on heat recirculation model in the proposed algorithms.

## References

[1] A. Beloglazov, J. Abawajy, R. Buyya, Energy-aware resource allocation heuristics for efficient management of data centers for Cloud computing, Future Gener. Comput. Syst., 28 (5), 2012, pp. 755-768.

[2] C.-C. Lin, P. Liu, J.-J. Wu, Energy-aware virtual machine dynamic provision and scheduling for cloud computing, in: Proceedings of the 2011 IEEE 4th International Conference on Cloud Computing, 2011.

[3] X. Wang, Z. Liu, An energy-aware VMs placement algorithm in cloud computing environment, in: Proceedings of the Second International Conference on Intelligent System Design and Engineering Application, 2012.

[4] J.T. Piao, J. Yan, A network-aware virtual machine placement and migration approach in cloud computing, in: Proceedings of the Ninth International Conference on Grid and Cloud Computing, 2010.

[5] W. Bing, L. Chuang, K. Xiangzhen, Energy optimized modeling for live migration in virtual data center, in Editor: Book Energy optimized modeling for live migration in virtual data center, 2011 edition, pp. 2311-2315.

[6] C. Clark, K. Fraser, S. Hand, J.G. Hansen, E. Jul, C. Limpach, I. Pratt, A. Warfield, Live migration of virtual machines, in: Proceedings of the 2nd conference on Symposium on Networked Systems Design & Implementation, Vol. 2, 2005.

[7] R. Raghavendra, P. Ranganathan, V. Talwar, Z. Wang, X. Zhu, No "power" struggles: coordinated multi-level power management for the data center, SIGARCH Comput. Archit. News, 36 (1), 2008, pp. 48-59.

[8] A. Verma, P. Ahuja, A. Neogi, pMapper: power and migration cost aware application placement in virtualized systems, in: Proceedings of the 9th ACM/IFIP/USENIX International Conference on Middleware, Leuven, Belgium, 2008.

[9] A. Gandhi, M. Harchol-Balter, R. Das, C. Lefurgy, Optimal power allocation in server farms, in: Proceedings of the eleventh international joint conference on Measurement and modeling of computer systems, Seattle, WA, USA, 2009.

[10] D. Kusic, J.O. Kephart, J.E. Hanson, N. Kandasamy, G. Jiang, Power and performance management of virtualized computing environments via lookahead control, Cluster Computing, 12 (1), 2009, pp. 1-15.

[11] HP Network Equipment Specifications, available online at http://h10010.www1.hp.com.

[12] E. Pakbaznia, M. Ghasemazar, M. Pedram, Temperature aware dynamic resource provisioning in a power optimized datacenter, in: Design, Automation and Test in Europe Conference and Exhibition (DATE10), 2010, pp. 124-129.

[13] L. Barroso, U. Holzle, The case for energy-proportional computing, Journal of IEEE Computer, January 2007.

[14] W. Voorsluys, J. Broberg, S. Venugopal, R. Buyya, Cost of virtual machine live migration in clouds: a performance evaluation, in: Proceedings of the 1st International Conference on Cloud Computing, CloudCom 2009, Springer, Beijing, China, 2009.

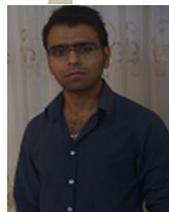
Sina Esfandiarpoor is an M.S. graduate from Computer Engineering Department of Sharif University of Technology, Kish campus, IRAN. He had joined Energy-Aware System Laboratory (EASY Lab) in 2011 to work towards his M.S. degree under supervision of Dr. Maziar Goudarzi. He has received his B.S. degree in Computer Engineering from Bahonar University of Kerman in 2009 and his M.S. in IT Communication Network from Sharif University of Technology in Kish campus in 2013. His research interests focus on enery-aware techniques specifically in the area of cloud computing.

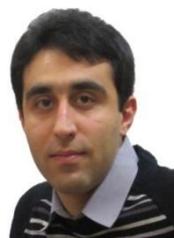
Ali Pahlavan is an M.S. graduate from Computer Engineering Department of Sharif University of Technology, IRAN. He had joined Energy-Aware System Laboratory (EASY Lab) in 2010 to work towards his M.S. degree under supervision of Dr. Maziar Goudarzi. He has received his B.S. degree in Computer Engineering from Ferdowsi University of Mashhad in 2010, and his M.S. in Computer Architecture from Sharif University of Technology in 2012. His research interests focus on system level energy optimization techniques specifically in the area of datacenters and cloud computing.

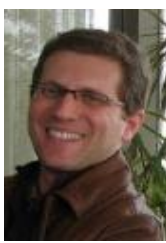
Maziar Goudarzi is an Assistant Professor at the Department of Computer Engineering, Sharif University of Technology, Tehran, Iran. He received the B.Sc., M.Sc., and Ph.D. degrees in Computer Engineering from Sharif University of Technology in 1996, 1998, and 2005, respectively. Before joining Sharif

University of Technology as a faculty member in September 2009, he was a Research Associate Professor at Kyushu University, Japan from 2006 to 2008, and then a member of research staff at University College Cork, Ireland in 2009. His current research interests include green computing, hardware–software codesign, and reconfigurable computing. Dr. Goudarzi has won two best paper awards, published several papers in reputable conferences and journals, and served as member of technical program committees of a number of IEEE, ACM, and IFIP conferences including ICCD, ASP-DAC, ISQED, ASQED, EUC, and IEDEC among others.